Title:
Symmetries by base substitutions in the genetic code predict 2' or 3' aminoacylation of tRNAs.

Authors: Jean-Luc Jestin[a], Christophe Soulé[b]

Addresses:

[a]Unité de Chimie Organique, URA 2128 CNRS
Département de Biologie Structurale et Chimie, Institut Pasteur
25 rue du Dr. Roux, 75724 Paris 15, France
email: jjestin@pasteur.fr (corresponding author)
tel +33 1 4438 9496; fax +33 1 4568 8404

[b]Institut des Hautes Etudes Scientifiques, CNRS
35 route de Chartres, 91440 Bures-sur-Yvette, France
email: soule@ihes.fr

Understanding why the genetic code is the way it is, has been the subject of numerous models and still remains largely a challenge (Freeland et al., 2000; Sella and Ardell, 2006). Associations between codons and amino acids were suggested to rely on RNA-amino acid interactions (Raszka and Mandel, 1972; Yarus, 1998). Closely related codons were put in correspondence with closely related amino acids within their biosynthetic pathways (Wong, 2005). Codons have also been grouped into systems characterized by interlocked thermodynamic cycles (Klump, 2006). Evolutionary models that minimise the number of the most frequent mutations provide a rationale for the fact that transitions at the third base of codons are mostly neutral mutations (Goldberg and Wittes, 1966). Similarly, minimization of the deleterious effects of sequence-dependent single-base deletions catalyzed by DNA polymerases provides a rationale for the assignment of stop signals to codons (Jestin and Kempf, 1997). While in-frame stop codons are strictly selected against, out-of-frame stop codons minimize the costs of ribosomal slippages (Seligmann and Pollock, 2004). In this context, the frequencies of codons were found to be highly dependent on the reading frame and highlighted a symmetrical codon pattern (Koch and Lehmann, 1997). As the genetic code is quasi-universal among living organisms, models do not need to be time-dependent, even though time-dependent models have been suggested (Bahi and Michel, 2004; Rodin and Rodin, 2006; Sella and Ardell, 2006). Symmetries in the genetic code are of special interest as they may highlight underlying organization principles of the code. A supersymmetric model for the evolution of the genetic code was proposed: successive breaking of these symmetries would provide an evolutive scenario for the decomposition into sets of synonymous codons (Hornos and Hornos, 1993; Bashford et al., 1997). When the amino acids are mapped to the vertices of a 28-gon, three two-fold symmetries were identified for three subsets of the cognate aminoacyl-tRNA synthetases (Yang, 2004).

This letter reports complete sets of two-fold symmetries between partitions of the universal genetic code. By substituting bases at each position of the codons according to a fixed rule, it happens that properties of the degeneracy pattern or of tRNA aminoacylation specificity are exchanged.

First the set of sixty-four codons of the genetic code was partitionned in two groups of thirty-two codons depending on whether the third base of triplets is necessary or not to define unambiguously an amino acid or a stop signal (property 1). Rumer reported a symmetry by base substitutions that alters property 1 (Rumer, 1966) . The substitutions exchanging T and G as well as A and C are applied to all three codon bases and are called Rumer's transformation. If the third base is necessary to define an amino acid, then the symmetrical codon by Rumer's transformation does not require the third base of codons to be defined so as to define unambiguously the amino acid. Conversely, if the third base does not have to be defined so as to define unambiguously an amino acid, then the symmetrical codon by Rumer's transformation requires the third base to be given so as to define unambiguously the amino acid. More recently, one of the authors reported a symmetry that leaves unchanged property 1 (Jestin, 2006): this symmetry consists in applying to the first base of codons the substitutions exchanging G and C as well as T and A. For example, GCN codons coding for alanine are exchanged into CCN codons coding for proline; for GCN and CCN codons, the third base does not have to be defined so as to define unambiguously the amino acid.

Here we report a third symmetry that alters property 1 (Fig.1). This symmetry is obtained by applying successively the two symmetries described above. It consists in applying the substitution exchanging A and G as well as C and T (a transition) to the first base in the codon, the substitution exchanging A and C as well as G and T (a transversion) to the second base in the codon, and the substitution exchanging A and C as well as G and T (a transversion) in the third base of the codon.

We show further that the only other symmetries exchanging both groups into each other are obtained by combining the previous ones with a symmetry acting only on the third base of the codons (here we do not include the substitution on the second base which exchanges A and C when fixing G and T). This can be seen by counting the number of occurrences of A, C, G, and T as first, second or third base in a codon of each group. The result is given in Table 1.

These symmetries are valid for the standard genetic code and for other genetic codes such as the vertebrate mitochondrial genetic code which has a higher degree of symmetry of its degeneracy pattern as noted earlier (Lehmann, 2000; Jestin, 2006).

In addition to the existence of Rumer's transformation, Shcherbak discussed the following Rumer's rule (Shcherbak, 1989), which can be read off Table 1: the ratio $R = C+G/T+A$ of the number of occurrences of C and G by the number of occurrences of T and A in positions 1, 2 and 3 is equal to 3, 3 and 1 respectively in codons of the first group (and hence it is 1/3, 1/3 and 1 for codons of the second group). Similarly, the ratio $P = T+C/A+G$ is 1, 3 and 1 in positions 1, 2 and 3 of the first group of codons.

Secondly, we considered another grouping of codons of the genetic code depending on whether the amino acids are acylated by amino acyl-tRNA synthetases at the 2' or at the 3' hydroxyl group of the tRNA's last ribose (property 2) (Sprinzl and Cramer, 1975; Arnez and Moras, 1994). This classification of amino acyl-tRNA synthetases is very similar to the one based on sequence homology and on structural considerations (Eriani et al., 1990; Cusack, 1997). Class I synthetases contain HIGH and KMSKS consensus sequences, which are absent from class II amino acyl tRNA synthetases. At the structural level, class I synthetases also contain a Rossman fold, a domain that binds nucleotides, unlike class II synthetases. Class I enzymes catalyse acylation at the 2' hydroxyl group of the tRNA while class II enzymes generally catalyse acylation at the 3' hydroxyl group of the tRNA. PheRS as a class II enzyme that catalyses acylation at the tRNA's 2' hydroxyl group is therefore an exception.

The case of cysteinyl-tRNA$^{Cys}$ synthetase (CysRS) is ambiguous and was investigated recently. CysRS is a class I synthetase, but establishes contacts with the major groove of the acceptor stem of the tRNA$^{Cys}$ as commonly found for class II enzymes. The enzyme from *Escherichia coli* is able to catalyse the acylation reaction at both 2' and 3' hydroxyl groups of the tRNA$^{Cys}$. The 2' acylation is about one order of magnitude faster than the 3' acylation when catalysed by *E. coli* cysteinyl-tRNA synthetase *in vitro* (Shitivelband and Hou, 2005).

The following classification was then used for 2' acylated amino acids (Ile, Leu, Met, Val, Trp, Tyr, Arg, Gln, Glu, Phe) and for 3' acylated amino acids (His, Pro, Ser, Thr, Asn, Asp, Lys, Ala, Gly). To the class of 2' acylated amino acids we also added the stop signals, a choice partially justified by the fact that two stop codons of the mitochondrial code of vertebrates code for the 2' acylated amino acid Arg in the universal code. Note that if cysteine were not in the class 3', or if a stop signal was not in the class 2', symmetries could not be identified. If cysteine is assigned to the class 2' as suggested by the previous paragraph, the symmetries are broken. Loss of the symmetries might have occurred during the evolution of aminoacyl-tRNA synthetases and might be associated to the late appearance of this amino acid in the genetic code (Brooks and Fresco, 2002).

When considering molecular properties such as polarity, volume and hydrophobicity, no statistical differences were noted between class 2' and class I on one hand, class 3' and class II on the other hand (Table 3).

There exist two symmetries by base substitutions that exchange the class 2' with the class 3' of the corresponding codon groups (cf. Fig.2). They consist in applying the substitution exchanging A and C as well as G and T (a transversion) to the first base of the codon, the substitution exchanging A and G as well as C and T (a transition) to the second base of the codon, and the substitution exchanging A and C as well as G and T or A and T as well as C and G (a transversion) to the third base of the codon. These two symmetries differ by the substitution exchanging A and G as well as C and T in the third position. They are not related to those depicted in Figures 4 and 5 (Yang, 2004) as Yang's three symmetries act only on three subsets of amino acids whereas the symmetries described herein are valid for the whole codon table.

There are no other symmetries by base substitutions between the two classes 2' and 3', as can be seen by counting the occurrences of A, C, G and T in each class and each position (Table 2). Note also the following analog of the Rumer's rule: both the ratio R

= C+G / T+A and the ratio Q = A+C / G+T are equal to 1, 1/3, 1 in positions 1, 2, 3 respectively in the class 2' (and 1, 3, 1 in the class 3').

In this letter we have described new symmetries by base substitutions in the genetic code for partitions concerning the codon degeneracy level or the tRNA-aminoacylation class. Several evolutionary models have been proposed concerning tRNAs and their aminoacyl-tRNA synthetases (Martinez Gimenez and Tabares Seisdedos, 2002; Klipcan and Safro, 2004; Chechetkin, 2006; Di Giulio, 2006). Newly introduced amino acids may well have been selected to minimize the deleterious effects of mistranslations, and possibly according to their molecular volumes (Torabi et al., 2006). A unique serie of binary divisions of the codon table was recently noted: when the same differentiation rule was applied at each division, the class I / class II pattern arose consistently (Delarue, 2007). Aminoacyl-tRNA synthetases are likely to have evolved by gene duplication and mutation of primordial synthetases within each class, as evidenced by sequence homology (Woese et al., 2000). Consistently, the symmetries highlighted in this manuscript require three base substitutions per codon, which are unlikely to happen, thereby shedding some light on the duplication and divergence mechanism of evolution among the two classes of aminoacyl-tRNA synthetases.

Acknowledgements :
We thank H. Epstein, E. Yeramian, D. Moras, B. Prum and J. Perona for their help.

8References :

Arnez, J. G., Moras, D. 1994. Aminoacyl-tRNA synthetase tRNA recognition. Oxford, IRL Press 61-81.

Bahi, J. M., Michel, C. J. 2004. A stochastic gene evolution model with time dependent mutations. Bull. Math. Biol. 66, 763-778.

Bashford, J. D., Tsohantjis, I., Jarvis, P. D. 1997. Codon and nucleotide assignments in a supersymmetric model of the genetic code. Phys. Lett. A 233, 481-488.

Brooks, D. J., Fresco, J. R. 2002. Increased frequency of cysteine, tyrosine, and phenylalanine residues since the last universal ancestor. Mol. Cell. Proteomics 1, 125-131.

Chechetkin, V. R. 2006. Genetic code from tRNA point of view. J. Theor. Biol. 242, 922-934.

Cusack, S. 1997. Aminoacyl-tRNA synthetases. Curr. Opin. Struct. Biol. 7, 881-889.

Delarue, M. 2007. An asymmetric underlying rule in the assignment of codons. RNA 13, 161-169.

Di Giulio, M. 2006. The non-monophyletic origin of the tRNA molecule and the origin of genes only after the evolutionary stage of the last universal common ancestor. J. Theor. Biol. 240, 343-352.

Di Giulio, M., Capobianco, M. R., Medugno, M. 1994. On the optimization of the physicochemical distances between amino acids in the evolution of the genetic code. J. Theor. Biol. 168, 43-51.

Eriani, G., Delarue, M., Poch, O., Gangloff, J., Moras, D. 1990. Partition of tRNA synthetases into two classes based on mutually exclusive sets of sequence motifs. Nature 347, 203-206.

Freeland, S. J., Knight, R. D., Landweber, L. F., Hurst, L. D. 2000. Early fixation of an optimal genetic code. Mol. Biol. Evol. 17, 511-518.

Figure Legends :

Figure 1

Exchange of Group I (codons for which the third base does not have to be defined to specify the amino acid) into Group II (codons for which the third base must be defined to specify unambiguously the amino acid or the stop signal) by the transformation (AG/CT for the first base, GT/AC for the second and third bases). N=A,T,G or C; H=A,T or C; Y=T or C; R=A or G.

Figure 2

Exchange of the classes 2' and 3' by the transformation (AC/GT on the first base, AG/CT on the second base, AC/GT on the third base). The special case of cysteine is labelled by an asterisk and discussed in the text.

Table I

Number of occurences of the bases A, C, G and T at each position within the codon in each group.

Table II

Number of occurences of the bases A, C, G and T at each position within the codon in each class.

Table III

Statistical t-values computed from the data on hydrophobicity (Kyte and Doolittle, 1982), molecular volume and polarity (Di Giulio et al., 1994) comparing the class 2' with class I, and the class 3' with class II. These values are below the threshold of significance given in the Student's table.

|  |  | A | C | G | T |
|---|---|---|---|---|---|
| Base 1 | Group I | 4 | 12 | 12 | 4 |
|  | Group II | 12 | 4 | 4 | 12 |
| Base 2 | Group I | 0 | 16 | 8 | 8 |
|  | Group II | 16 | 0 | 8 | 8 |
| Base 3 | Group I | 8 | 8 | 8 | 8 |
|  | Group II | 8 | 8 | 8 | 8 |

Table 1

|        |          | A  | C  | G  | T  |
|--------|----------|----|----|----|----|
| Base 1 | Class 2' | 6  | 10 | 6  | 10 |
|        | Class 3' | 10 | 6  | 10 | 6  |
| Base 2 | Class 2' | 8  | 0  | 8  | 16 |
|        | Class 3' | 8  | 16 | 8  | 0  |
| Base 3 | Class 2' | 10 | 6  | 10 | 6  |
|        | Class 3' | 6  | 10 | 6  | 10 |

Table 2

|  | Class 2' / Class I | Class 3' / Class II |
|---|---|---|
| Hydrophobicity | 0.07 | 0.11 |
| Polarity | 0.017 | 0.019 |
| Volume | 0.57 | 0.45 |

Table 3